# Engineering Interaction Dynamics in Active Resonant Photonic Structures

Yuzhou G. N. Liu[1], Omid Hemmatyar[1], Absar U. Hassan[2,3], Pawel S. Jung[2,4], Jae-Hyuck Choi[1], Demetrios N. Christodoulides[2], and Mercedeh Khajavikhan[1,2,a]

[1]*Ming Hsieh Department of Electrical and Computer Engineering, University of Southern California, Los Angeles, CA 90089, USA*

[2]*CREOL, The College of Optics and Photonics, University of Central Florida, Orlando, FL 32816-2700, USA*

[3]*Facebook inc., Menlo Park, CA 94025, USA*

[4]*Faculty of Physics, Warsaw University of Technology, Koszykowa 75, 00-662 Warsaw, Poland*

[a)]Author to whom correspondence should be addressed: khajavik@usc.edu

The collective response of a system is profoundly shaped by the interaction dynamics between its constituent elements. In physics, tailoring these interactions can enable the observation of unusual phenomena that are otherwise inaccessible in standard settings, ranging from the possibility of a Kramer's degeneracy even in the absence of spin to the breakdown of the bulk-boundary correspondence. Here, we show how such tailored asymmetric coupling terms can be realized in photonic integrated platforms by exploiting non-Hermitian concepts. In this regard, we introduce a generalized photonic molecule composed of a pair of microring resonators with internal S-bends connected via two directional couplers and a link waveguide. By judiciously designing the parameters of this system, namely the length of the links and the power division ratio of the directional couplers, we experimentally show the emergence of Hermitian and non-Hermitian type exchange interactions. The ramifications of such coupling dynamics are then studied in 1D chain and ring-type active lattices. Our findings establish the proposed structure as a promising building block for the realization of a variety of phenomena, especially those associated with phase locking in laser arrays and non-Hermitian topological lattices.

## I. INTRODUCTION

Synthetic gauge fields, enabling photons to flow in an elaborately designed photonic lattice in a similar fashion as electrons in a magnetic flux, are the cornerstone of photonic topological insulators. In optics, such fields have been primarily realized through geometrical design and time/space modulation[1–9]. Most early photonic topological lattices have been built on passive platforms, where the topological response can be examined by probing the system with an input and observing the resulting output[4–6]. However, it has been recently realized that such topological attributes may be more easily witnessed and consequently utilized in active settings under pumping, especially when the system reaches lasing. In such cases, instead of externally exciting the system, one can deduce the signatures of the underlying topology from the emission spectrum and intensity profile. Along these lines, topological lasers have been demonstrated in which an array of gain elements phase lock to yield a topological edge supermode[10–18]. Such lasers have been shown to possess superior properties when compared to their trivial counterparts, in terms of spectral and spatial purity of their emission, higher quantum efficiency and robustness to defects and disorders[12,16]. The use of gain in topological photonics, however, is of more significance than just simplifying measurements, or even enabling a new class of lasers. Using optical amplification, one can fundamentally change the nature of interactions between the resonant units in an array[19]. In other words, unlike passive systems, in which reciprocity requires the exchange dynamics between elements to be symmetrical, in active/non-Hermitian systems, one can engineer such couplings to be, in general, asymmetric.

This is because Lorentz reciprocity is defined in the context of source-free arrangements that are excited by an external source, while in active systems, the presence of gain saturation and spontaneous emission eludes a clear discussion of reciprocity in the abstract definition. In fact, it can be shown that spontaneous emission in judiciously designed resonant structures can lead to non-symmetric coupling interactions[19].

In this article, we propose and experimentally demonstrate a versatile approach for engineering the coupling dynamics in active cavities. In order to do so, we use microring resonators with internal S-bend constructs to enforce unidirectional lasing[12,20,21]. The coupling is then established using a link structure. By changing the length of the link and the power division factor of the directional couplers, various types of Hermitian and non-Hermitian coupling behaviors can be realized. The freedom to design at will the interaction dynamics in a lattice can bring many possibilities to optics and in particular to topological photonics[22–29]. Along these lines, some of the applications of such active lattices will be discussed.

The paper is organized as follows. In Section II, an active two-resonator system with an adjustable coupling link is introduced. We show how the coupling can be altered from Hermitian- to non-Hermitian-type exchange by varying the lengths and the power division ratio of the directional couplers. In Section III, the response of this class of coupled cavities will be verified experimentally. Section IV explores the ramifications of non-Hermitian coupling in 1D chain and ring arrays. Finally, in Section V we will conclude the paper.

## II. GENERALIZED PHOTONIC MOLECULE FOR NON-HERMITIAN COUPLING

Photonic molecules, representing two coupled resonant structures, have been studied extensively in the past few decades[30,31]. In standard optical systems that respect time reversal symmetry, the energy exchange between the constituent elements of a photonic molecule tend to be symmetric, resulting in a Hermitian type dynamical response. Here, we introduce a new type of active photonic molecule that can offer a variety of exchange interactions from Hermitian- to non-Hermitian-type through complex coupling terms. Figure 1a shows the schematic of such a molecule. This structure is composed of two active (displaying gain) ring resonators with internal S-bends, connected to each other through a combination of cascaded directional couplers (DCs) that are separated by a link waveguide. Here, the S-bends are designed in such a way to enforce unidirectional lasing in the rings[20] (clockwise for the case depicted in Fig. 1). The left (right) ports of the first (second) directional coupler are weakly coupled to the opposing sides of ring ①(②).

In order to analyze this system, we can write the spatial coupled mode equations relating the fields in the two rings- without a priori assumption regarding the unidirectional flow of light in the cavities (see Appendix A for details). The result, however, appears to fully agree with the temporal coupled mode analysis performed with the assumption that the two rings are unidirectional. Without loss of generality, here we use the latter approach. In addition, in order to keep the number of variables limited, we study the case where all coupling strengths between resonators and waveguides are equal and the two directional couplers are identical. In general, each ring, when uncoupled, can support a number of spectral lines. We also limit our analysis to a single longitudinal mode- since these modes can be treated independently. We also assume that the waveguiding section is designed so as to primarily support the $TE_0$ transverse mode. In this respect, the interplay between the electric modal fields in the two identical rings can be effectively described through a set of time-dependent coupled equations:

$$\begin{cases} i\dot{a}_1 + \omega_0 a_1 + \kappa_{2\to1} a_2 = 0 \\ i\dot{a}_2 + \omega_0 a_2 + \kappa_{1\to2} a_1 = 0 \end{cases} \quad (1)$$

where $a_{1,2}$ represent the modal amplitudes in the two cavities. The angular frequency $\omega_0$ is determined by the resonance conditions for each resonator in the absence of coupling. The coupling from resonator ① → ② is given by $\kappa_{1\to2} = i\kappa^2 r^2 e^{i\beta L}$, while that from ② → ① is expressed by $\kappa_{2\to1} = -i\kappa^2 t^2 e^{i\beta L}$ (please see *Appendix B* for the derivation). Here, $r$ and $t$ are the through and cross terms for the directional coupler. In a lossless device, these two coefficients are related through $r^2 + t^2 = 1$. The coupling strengths between the rings and bus waveguides are all set to be $\kappa$. Finally, the overall length of the link part is given by $L = L_0 + 2L_c$, where $L_0$ and $L_c$ are the lengths of the waveguide sections between the two directional couplers and from each directional coupler to the adjacent ring, respectively.

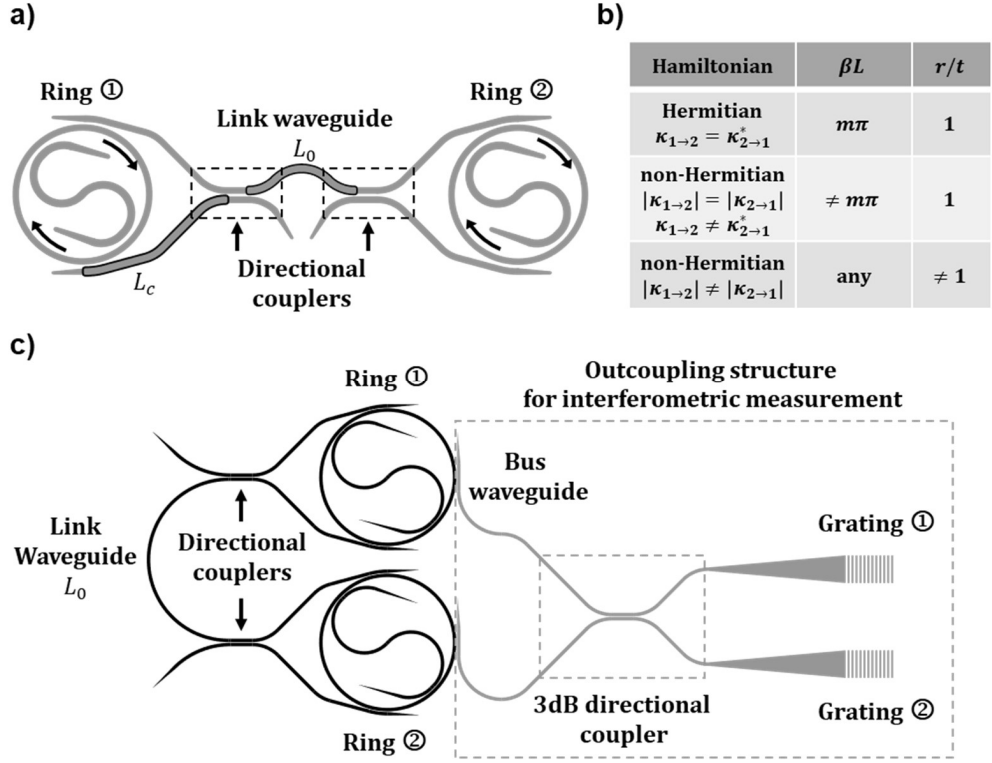

**FIG. 1.** (a) Schematic of a two-element system with unidirectional microring resonators and a link structure. The cascaded directional couplers provide a $\pi$ phase difference between the two coupling directions. The coupling phase $\beta L$ can be changed by altering the length $L$. An asymmetric coupling coefficient can be introduced by selecting the length of the directional couplers. (b) Three types of couplings under different settings of $\beta L$ and $r/t$ can be achieved. (c) A two-element system with an outcoupling structure for device characterization.

Depending on the values of $\beta L$ and $r$ (or equivalently $t$), the Hamiltonian describing this system can become Hermitian, non-Hermitian with two coupling coefficients of equal magnitudes but opposing signs, or non-Hermitian with interactions displaying dissimilar magnitude and/or phase. For example, in the case of $r = t$ (3dB directional couplers), when $\beta L = m\pi$, this system exhibits a fully Hermitian behavior, while for $\beta L \neq m\pi$, the coupling between the elements and therefore the Hamiltonian becomes non-Hermitian, even though the magnitude of the exchange interactions remains the same. On the other hand, for $r \neq t$, the Hamiltonian is inevitably non-Hermitian since the magnitude of the coupling terms will be different. If non-Hermitian interactions that are purely imaginary are desired, one can change the length $L_c$ of the upper arms vs. those on the lower arms. The table provided in Fig. 1(b) summarizes the conditions leading to the aforementioned various scenarios. It should be noted that even though the non-Hermitian coupling here is enabled by gain, the type of non-Hermiticity we observe in these systems is very different from those observed in PT-symmetric arrangements that are realized by the presence of a gain-loss contrast[32–35].

## III. EXPERIMENTAL OBSERVATION OF GENERALIZED COUPLING INTERACTIONS IN 2-ELEMENT SYSTEMS

In order to verify the response of the generalized photonic molecule introduced in the previous section, here we fabricate the proposed resonant systems on a III-V semiconductor wafer (6 quantum wells of InGaAsP with an overall thickness of 200 nm). The ring resonators and all waveguides have a width of 500 nm and a height of 200 nm. The high-contrast core ($n_{core} = 3.4$) is embedded in silicon dioxide ($n_{SiO_2} = 1.45$) and is exposed to air on top. This structure is designed to support the $TE_0$ mode with an effective index of $n_{eff} = 2.273$ and a group index of $n_g \approx 4$ at a wavelength of 1580 nm. The separations between the waveguides in the directional couplers, the bus WGs and resonators, as well as the S-bends and resonators are all nominally set to be 100 nm.

By providing adequate pumping to this system, the eigenmodes with the larger imaginary part of the eigenvalues (gain) are expected to lase, allowing the emission to display the signatures of the underlying coupling. In particular, one can measure the spectrum as well as the phase difference between the outputs from the two resonators. For measuring the phase, we collect the light from the two resonators with bus waveguides and allow them to interfere using an additional 3dB coupler as depicted in Fig. 1(c). The output arms of the directional coupler are equipped with surface emitting second-order gratings.

Figure 2 depicts three different scenarios of interest, where the coupling turns from non-Hermitian to Hermitian and back to non-Hermitian again, simply by changing the length of the link $L_0$. When $\beta L = 2m\pi + \pi/2$, the coupling terms ($\kappa_{1\to 2}$ and $\kappa_{2\to 1}$) are both real but with opposing signs. The non-Hermitian coupling allows the system to support two modes with eigenvectors $[1 \quad \pm i]^T$ and corresponding eigenvalues $\mp i|\kappa_{1\to 2}|$. This presents a situation where the structure supports two modes with equal field intensity in resonators ① and ②. The electric field components from the mode with the higher quality factor (subject to gain) show a $-\pi/2$ phase difference, while this quantity for a mode with a lower quality factor (experiencing loss) is $+\pi/2$. Figure 2(a) shows the lasing properties of such a photonic molecule. The emission intensity clearly corroborates the expected interference behavior, as the light exclusively exits the bottom grating port. In addition, the spectrum shows a single peak emission, an indication that one of the modes is largely suppressed and the laser is now single-moded.

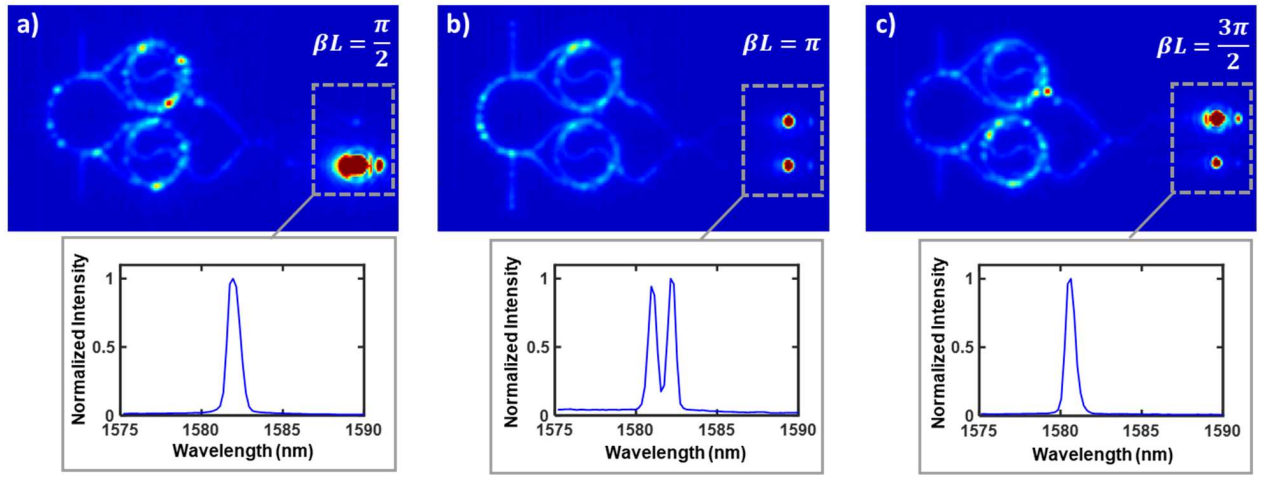

**FIG. 2.** Experimental results of coupling coefficients for three values of $\beta L$, (a) $\beta L = 2m\pi + \pi/2$, (b) $\beta L = m\pi$, and (c) $\beta L = 2m\pi - \pi/2$. Insets show the spectra collected at the output gratings. The spectrum in (a) and (c) indicate that there is only one lasing mode when a non-Hermitian coupling is implemented. The spectrum in (b) shows there are two lasing modes when the two resonators interact through a Hermitian coupling.

The above photonic molecule, however, exhibits a very different kind of response when the length of link structure satisfies $\beta L = m\pi$. In this case, the two coupling terms become imaginary and complex conjugate of each other. The resulting Hermitian Hamiltonian gives rise to two eigenmodes of $[1 \quad \pm i]^T$ with eigenvalues of $\pm|\kappa_{1\to 2}|$. In the absence of mode discrimination, this system is expected to support two lasing modes that are separated by a frequency difference of $2|\kappa_{1\to 2}|$. This situation can be clearly seen in Fig. 2(b) where not only both grating ports have a balanced output intensity, but also the spectrum shows two lasing lines. Finally, Fig. 2(c) shows a similar Hermitian scenario as in Fig. 2(a), albeit with $\beta L = 2m\pi - \pi/2$. In this case, the system is again single-moded, but the two components of the electric field associated with the lasing mode are displaying a $+\pi/2$ phase difference, resulting in a higher output power emerging from the upper grating.

## V. GENERALIZED INTERACTIONS IN 1D CHAIN AND RING-SHAPED LATTICES

The nature and strength of the interconnectivity between the elements of a lattice can profoundly affect its band structure. In this section, we study the ramifications of generalized coupling dynamics on the band shape, energy states, modal content, and particularly the emergence of edge modes in lattices. In this regard, we consider several 1D chains in which the elements are coupled through imaginary Hermitian and non-Hermitian interactions. Subsequently, we extend our analysis to the ring-type lattice configurations.

Generally, the eigenvalues of a 1D chain of length N with nearest neighbor right-to-left ($\kappa_{r\to l}$) and left-to-right $\kappa_{l\to r}$ couplings, as depicted in Fig. 3(a), are given by:

$$\lambda_h = \omega_0 - 2\sqrt{\kappa_{l\to r}\kappa_{r\to l}} \cos\left(\frac{h\pi}{N+1}\right)$$

Here, $h$ is the integer identifying the mode number ($h = 1:N$). The corresponding eigenvectors are then given by:

$$V_h = [v_{h,1}, v_{h,2}, \dots, v_{h,N}]^T$$

Where $v_{h,k} = (\kappa_{l\to r}/\kappa_{r\to l})^{k/2} \sin\left(\frac{hk\pi}{N+1}\right); \; k = 1:N$.

Figure 3(b) shows an implementation of such a 1D lattice using our generalized photonic molecules as the constituent elements. When the coupling is designed to be purely imaginary, but the Hamiltonian is Hermitian (all $L_{c_{i,j}}$s are equal, and $\beta L$ is a multiple integer of $2\pi$ resulting in $i\kappa_{l\to r} = -i\kappa_{r\to l}$); all eigenvalues are expected to be real, thus representing modes that are separated in the spectral domain while having the same quality factors. Figure 3(c) shows the eigen-spectrum as well as the fundamental mode of this system. On the other hand, when the coupling terms are purely imaginary but equal in magnitude ( $L_{c_{i,1}} = L_{c_{i,2}} = L_{c_{i+1,1}} = 2q\pi/\beta$, and $L_{c_{i+1,2}} = L_{c_{i+2,1}} = (2q-1)\pi/\beta$, and $\beta L_0 = 2m\pi$ ), resulting in a Hamiltonian that is non-Hermitian ($i\kappa_{l\to r} = i\kappa_{r\to l}$); all the eigenvalues become entirely imaginary with the in-phase mode experiencing a net gain (positive imaginary term), while the out-of-phase mode undergoes the same amount of loss. The eigenvalue distribution as well as the amplitude of the fundamental mode of this arrangement are depicted in Fig. 3(d). This situation may be of interest in laser phase locking where the mode discrimination offered by the difference between the eigenvalues forces the array into lasing in the in-phase supermode. This scheme may be of practical use in fiber laser systems where one can control the locking dynamics against the random variations of the individual cavity lengths by adjusting the coupling strength. It should be noted that the intensity distribution across the various elements of the lattice can be further evened out by slightly changing the coupling strengths between those elements of the chain that are closer to the ends.

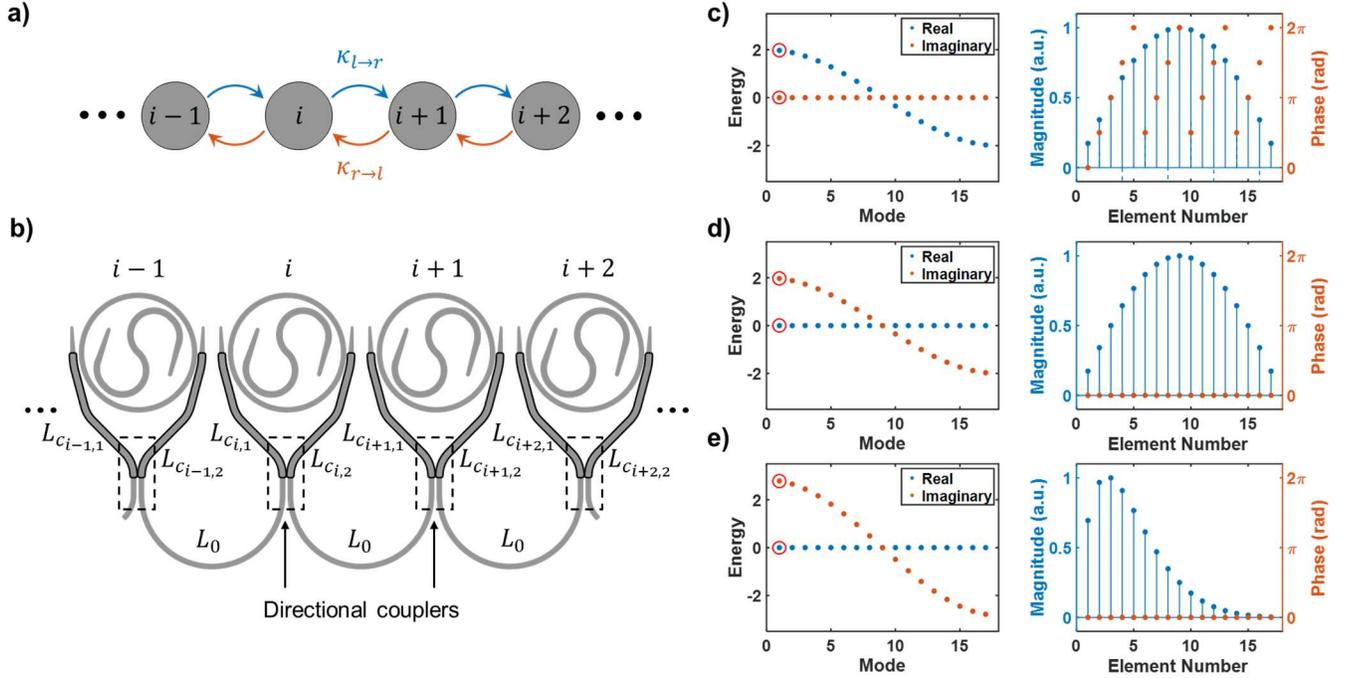

**FIG. 3.** (a) A 1D optical chain with asymmetric couplings. (b) The photonic realization of a chain where the of length of bus waveguides $L_{c_{i,j}}$, $L_0$, and the coupling ratios of the directional couplers ($r/t$) can be modified to realize various types of interactions. Eigenvalues associated with different modes, as well as the amplitude of the components of the fundamental mode for chains with (c) imaginary Hermitian coupling ($\kappa_{l \to r} = -\kappa_{r \to l} = i\kappa^2/2$), (d) balanced imaginary non-Hermitian coupling ($\kappa_{l \to r} = \kappa_{r \to l} = i\kappa^2/2$), and (e) unbalanced imaginary non-Hermitian coupling ($2\kappa_{l \to r} = \kappa_{r \to l} = i\kappa^2$) resulting in a non-Hermitian topological lattice.

Finally, when the coupling between the array elements become imaginary and uneven (all $L_{c_{i,j}}$s are equal, $\beta L = 2m\pi$, and $r/t \neq 1$), a new type of topological edge mode appears in these lattices. This situation known as Hatano-Nelson model was first proposed to describe the localization transition in systems with non-Hermitian couplings $|\kappa_{l \to r}| \neq |\kappa_{r \to l}|$[26,27]. Despite a number of proposals, the implementation of this lattice has yet remained elusive in optics. Using our generalized photonic molecules, the Hatano-Nelson chain can be realized by modifying the directional couplers in the lattice shown in Fig. 3(b) to have unequal through and cross coupling terms ($r \neq t$). When a chain structure is formed based on these molecules, all eigenvalues split in the imaginary part of the frequency domain, while the eigenvectors/field distributions become unbalanced, tilting towards one end of lattice (Fig. 3(e)). Unlike the standard Su-Schrieffer-Heeger (SSH) arrays[36], in this case, all modes of the system are of edge-type, leading to a situation that is coined by bulk-boundary correspondence[28,37]. The appearance of edge-type modes in this system should be of no surprise since the left-right coupling unbalance pushes the energy towards one end of the array. In such arrangements, the edge mode that represents the closest to a uniform distribution of field intensity at various elements experiences the highest level of gain.

Ring-type lattices offering periodic boundary conditions in addition to their inter-element spacing are used to study a broad range of physical phenomena from spin-squeezing to various topological and synthetic gauge structures. Figure 4(a) shows a schematic of a 4-element ring-shaped lattice where the left-to-right coupling differs from right-to-left exchange. Unlike the 1D chains, the circulation of power in such toroidal lattices leads to the formation of modes with equal intensities across the array elements. In such systems, a Hermitian imaginary coupling $i\kappa_{l \to r} = -i\kappa_{r \to l}$ (that is realized by having all $L_{c_{i,j}}$s to be equal, and $\beta L$ to be a multiple integer of $2\pi$ in Fig. 4(b)) does not provide any mode selectivity, but it gives rise to a fundamental mode with equal intensity in all sites (see Fig. 4(c)). On the other hand, when the coupling between elements is of non-Hermitian and imaginary type while the magnitudes of the coupling terms are equal ($|\kappa_{1\to 2}| = |\kappa_{2\to 1}|$) (for example by choosing $L_{ci,1} = L_{ci,2} = L_{ci+1,1} = 2q\pi/\beta$, and $L_{ci+1,2} = L_{ci+2,1} = (2q-1)\pi/\beta$, and $\beta L_0 = 2m\pi$), the eigenfrequencies exhibit a splitting in the complex domain, resulting in a mode discrimination (Fig. 4(d)), where the in-phase super-mode exhibits the highest quality factor. Finally, when the coupling is non-Hermitian and imaginary and $|\kappa_{1\to 2}| \neq |\kappa_{2\to 1}|$ (by having all $L_{ci,j}$s to be equal, $\beta L = 2m\pi$, and $r/t \neq 1$), mode discrimination can be observed both in real and imaginary parts of the eigenfrequency (Fig. 4(e)). As a result, various modes not only differ in their quality factors, but appear at different parts of the spectrum. Here, no apparent edge mode can be identified, even though the modes show natural robustness against the variation in couplings between the elements.

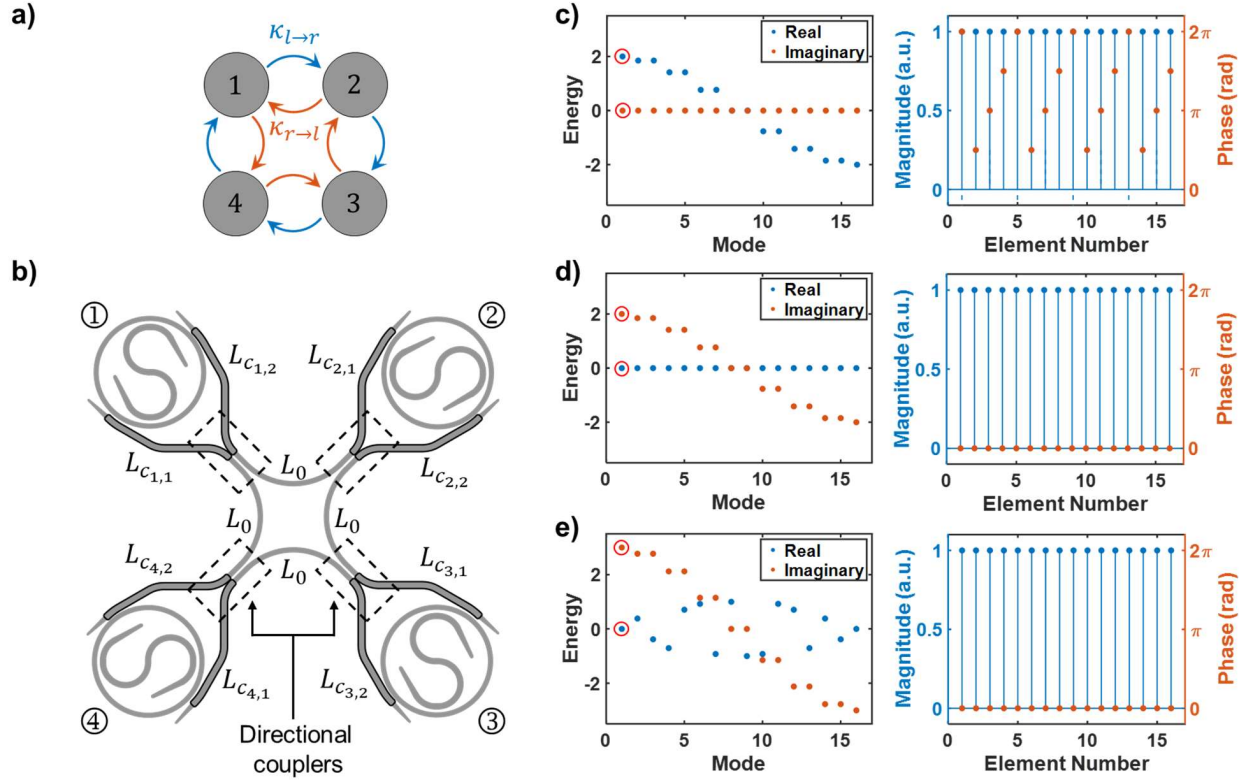

**FIG. 4.** (a) Schematic of a 4-element loop-type lattice with dissimilar $\kappa_{l\to r}$ and $\kappa_{r\to l}$ coupling terms. (b) A photonic realization of a 4-element loop. To implement different types of couplings, one can engineer the of length of the bus waveguides $L_{ci,j}$, $L_0$, and the coupling ratios of the directional couplers ($r/t$) resulting in (c) imaginary Hermitian coupling ($\kappa_{l\to r} = -\kappa_{r\to l} = i\kappa^2/2$), (d) balanced non-Hermitian coupling ($\kappa_{l\to r} = \kappa_{r\to l} = i\kappa^2/2$), and (e) unbalanced imaginary non-Hermitian coupling ($2\kappa_{l\to r} = \kappa_{r\to l} = i\kappa^2$), respectively. Eigenenergy corresponding to the various modes, as well as the amplitude of the components of the fundamental mode for loops are plotted. With imaginary coupling coefficients, the field profiles are found to be uniform. The fundamental modes in (d) and (e) have the highest quality factors.

## VI. CONCLUSION

We introduced a new active photonic molecule capable of exhibiting a wide range of interaction dynamics by adjusting its parameters. What enables this system to work, is the unidirectional flow of energy in the rings due to the presence of the S-bends. The emergence of various types of coupling coefficients in this molecule has been experimentally verified resulting in

Hermitian and non-Hermitian-type Hamiltonians. The exchange terms can be, in general, complex, even though in most of our analysis we focused on purely imaginary couplings. We have also explored the role these interaction dynamics play in determining the response of 1D chains and ring-shaped lattices. This building block may be used to study a variety of effects, especially those arising in non-Hermitian topological lattices. In addition, engineering the interaction dynamics in laser arrays comprising of such molecules may lead to new regimes of phase locking as well as near-field and far-field steerable emission.

## ACKNOWLEDGEMENT

The authors would like to thank the support from DARPA (D18AP00058), Office of Naval Research (N00014-20-1-2522, N00014-20-1-2789, N00014-16-1-2640, N00014-18-1-2347, N00014-19-1-2052), Army Research Office (W911NF-17-1-0481), Air Force Office of Scientific Research (FA9550-14-1-0037, FA9550-20-1-0322), National Science Foundation (CBET 1805200, ECCS 2000538, ECCS 2011171), US–Israel Binational Science Foundation (BSF; 2016381), and Polish Ministry of Science and Higher Education (Mobility Plus, 1654/MOB/V/2017).

## DATA AVAILABILITY

The data that support the findings of this study are available from the corresponding author upon reasonable request.

## APPENDICES

### A. Spatial coupled mode analysis of the generalized photonic molecule

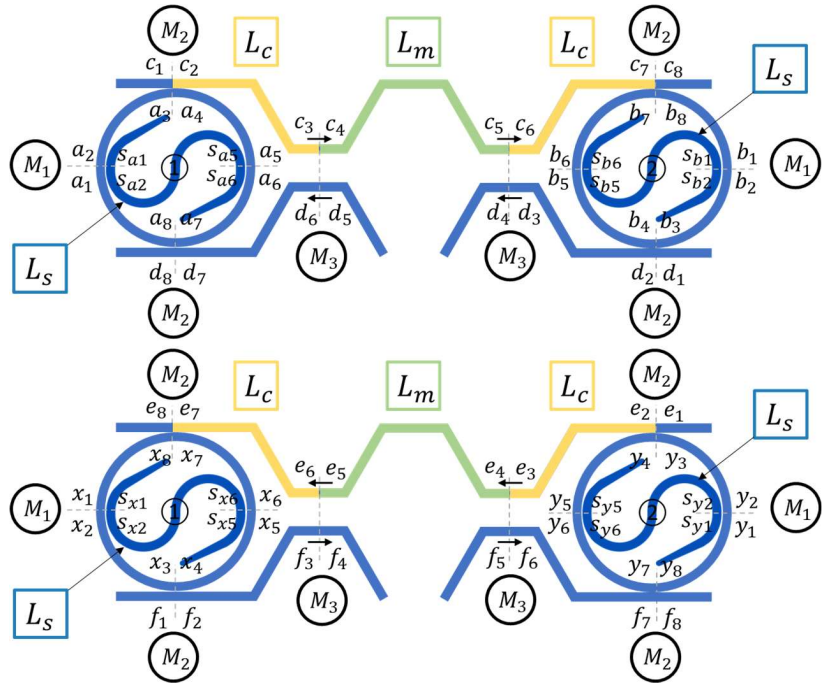

**FIG. 5.** Two microring resonators coupled through a link involving two directional couplers. Consider fields with opposite directions.

To derive the aforementioned coupling coefficients with spatial coupled mode analysis, we introduce the transfer matrices associated with the various elements involved in the links between the two resonators as shown in Fig. 5. In particular,

$$M_1 = \begin{pmatrix} \sqrt{1-\epsilon_1^2} & i\epsilon_1 \\ i\epsilon_1 & \sqrt{1-\epsilon_1^2} \end{pmatrix}, \quad M_2 = \begin{pmatrix} \sqrt{1-\epsilon_2^2} & i\epsilon_2 \\ i\epsilon_2 & \sqrt{1-\epsilon_2^2} \end{pmatrix} \quad (A1)$$

$$M_3 = \begin{pmatrix} r & it \\ it & r \end{pmatrix} \quad (A2)$$

where $\sqrt{1-\epsilon_j^2}$ and $i\epsilon_j$ represent the portion of the field that propagates through and cross the coupling regions, respectively.

The $M_3$ matrices correspond to the two directional couplers, while the $M_1$ and $M_2$ matrices account for weak coupling to the S-bends and waveguide buses, respectively. Consider the fields propagating along the direction of the S-bend, i.e., counterclockwise in Fig. 5. The coupling between the S-bends and the resonators can be written as:

$$\begin{pmatrix}a_2\\s_{a2}\end{pmatrix} = M_1 \begin{pmatrix}a_1\\s_{a1}\end{pmatrix}, \quad \begin{pmatrix}a_6\\s_{a6}\end{pmatrix} = M_1 \begin{pmatrix}a_5\\s_{a5}\end{pmatrix}, \quad \begin{pmatrix}b_2\\s_{b2}\end{pmatrix} = M_1 \begin{pmatrix}b_1\\s_{b1}\end{pmatrix}, \quad \begin{pmatrix}b_6\\s_{b6}\end{pmatrix} = M_1 \begin{pmatrix}b_5\\s_{b5}\end{pmatrix}, \tag{A3}$$

and the coupling between the resonator and the links can be expressed via:

$$\begin{pmatrix}a_4\\c_2\end{pmatrix} = M_2 \begin{pmatrix}a_3\\c_1\end{pmatrix}, \quad \begin{pmatrix}a_8\\d_8\end{pmatrix} = M_2 \begin{pmatrix}a_7\\d_7\end{pmatrix}, \quad \begin{pmatrix}b_4\\d_2\end{pmatrix} = M_2 \begin{pmatrix}b_3\\d_1\end{pmatrix}, \quad \begin{pmatrix}b_8\\c_8\end{pmatrix} = M_2 \begin{pmatrix}b_7\\c_7\end{pmatrix}. \tag{A4}$$

On the other hand, consider the fields propagating against the direction of the S-bend, i.e., clockwise in Fig. 5. The coupling between the S-bends and the resonators can be written as:

$$\begin{pmatrix}x_2\\s_{x2}\end{pmatrix} = M_1 \begin{pmatrix}x_1\\s_{x1}\end{pmatrix}, \quad \begin{pmatrix}x_6\\s_{x6}\end{pmatrix} = M_1 \begin{pmatrix}x_5\\s_{x5}\end{pmatrix}, \quad \begin{pmatrix}y_2\\s_{y2}\end{pmatrix} = M_1 \begin{pmatrix}y_1\\s_{y1}\end{pmatrix}, \quad \begin{pmatrix}y_6\\s_{y6}\end{pmatrix} = M_1 \begin{pmatrix}y_5\\s_{y5}\end{pmatrix}, \tag{A5}$$

and the coupling between the resonator and the links can be expressed via:

$$\begin{pmatrix}x_4\\f_2\end{pmatrix} = M_2 \begin{pmatrix}x_3\\f_1\end{pmatrix}, \quad \begin{pmatrix}x_8\\e_8\end{pmatrix} = M_2 \begin{pmatrix}x_7\\e_7\end{pmatrix}, \quad \begin{pmatrix}y_4\\e_2\end{pmatrix} = M_2 \begin{pmatrix}y_3\\e_1\end{pmatrix}, \quad \begin{pmatrix}y_8\\f_8\end{pmatrix} = M_2 \begin{pmatrix}y_7\\f_7\end{pmatrix}. \tag{A6}$$

In addition, the field amplitudes at the directional couplers are

$$\begin{pmatrix}c_4\\f_4\end{pmatrix} = M_3 \begin{pmatrix}c_3\\f_3\end{pmatrix}, \quad \begin{pmatrix}c_6\\f_6\end{pmatrix} = M_3 \begin{pmatrix}c_5\\f_5\end{pmatrix}, \quad \begin{pmatrix}d_4\\e_4\end{pmatrix} = M_3 \begin{pmatrix}d_3\\e_3\end{pmatrix}, \quad \begin{pmatrix}d_6\\e_6\end{pmatrix} = M_3 \begin{pmatrix}d_5\\e_5\end{pmatrix}. \tag{A7}$$

Assuming $\alpha$ to be the amplifying/damping rate per unit length and $L_r$ to be the perimeter of each resonator. The field in resonator ① propagate according to

$$a_1 = e^{\frac{i\beta L_r}{4}-\frac{\alpha L_r}{4}}a_8, \quad a_3 = e^{\frac{i\beta L_r}{4}-\frac{\alpha L_r}{4}}a_2, \quad a_5 = e^{\frac{i\beta L_r}{4}-\frac{\alpha L_r}{4}}a_4, \quad a_7 = e^{\frac{i\beta L_r}{4}-\frac{\alpha L_r}{4}}a_6 \tag{A8}$$

Similarly, the field components inside resonator ② obey

$$b_1 = e^{\frac{i\beta L_r}{4}-\frac{\alpha L_r}{4}}b_8, \quad b_3 = e^{\frac{i\beta L_r}{4}-\frac{\alpha L_r}{4}}b_2, \quad b_5 = e^{\frac{i\beta L_r}{4}-\frac{\alpha L_r}{4}}b_4, \quad b_7 = e^{\frac{i\beta L_r}{4}-\frac{\alpha L_r}{4}}b_6 \tag{A9}$$

In addition, the length from the resonator to the directional coupler is $L_c$, the length between two directional couplers is $L_m$, and the length of the S-bend is $L_s$ as shown in Fig. 5. By neglecting any amplification/loss in the link, the fields propagate according to:

$$c_3 = e^{i\beta L_c}c_2, \quad c_5 = e^{i\beta L_m}c_4, \quad c_7 = e^{i\beta L_c}c_6 \tag{A10}$$
$$d_3 = e^{i\beta L_c}d_2, \quad\quad\quad\quad\quad\quad d_7 = e^{i\beta L_c}d_6 \tag{A11}$$
$$e_3 = e^{i\beta L_c}e_2, \quad e_5 = e^{i\beta L_m}e_4, \quad e_7 = e^{i\beta L_c}e_6 \tag{A12}$$
$$f_3 = e^{i\beta L_c}f_2, \quad\quad\quad\quad\quad\quad f_7 = e^{i\beta L_c}f_6 \tag{A13}$$
$$s_{a2} = e^{i\beta L_s}s_{x6}, \quad\quad\quad\quad\quad s_{a5} = e^{i\beta L_s}s_{x2} \tag{A14}$$
$$s_{b2} = e^{i\beta L_s}s_{y6}, \quad\quad\quad\quad\quad s_{b5} = e^{i\beta L_s}s_{y2} \tag{A15}$$

The transfer matrix of this system can be written as:

$$\begin{bmatrix} K_{aa} & K_{ba} & K_{xa} & K_{ya} \\ K_{ab} & K_{bb} & K_{xb} & K_{yb} \\ K_{ax} & K_{bx} & K_{xx} & K_{yx} \\ K_{ay} & K_{by} & K_{xy} & K_{yy} \end{bmatrix} \begin{bmatrix} a_1 \\ b_1 \\ x_1 \\ y_1 \end{bmatrix} = \widehat{K} \begin{bmatrix} a_1 \\ b_1 \\ x_1 \\ y_1 \end{bmatrix} = \Lambda \begin{bmatrix} a_1 \\ b_1 \\ x_1 \\ y_1 \end{bmatrix} \tag{A16}$$

where $\Lambda$ is the system eigenvalue when $[a_1 \quad b_1 \quad x_1 \quad y_1]^T$ represent the system eigenvectors. The elements $K_{mn}$ in the matrix $\widehat{K}$ are:

$$K_{aa} = e^{-L_r\alpha + iL_r\beta}(-1 + \epsilon_1^2)(-1 + \epsilon_2^2) \tag{A17}$$

$$K_{ba} = \frac{1}{2} e^{-\frac{1}{2}L_r(\alpha - i\beta) + i(2\beta L_c + \beta L_m)} \sqrt{1 - \epsilon_1^2} \epsilon_2^2 \tag{A18}$$

$$K_{xa} = -e^{-\frac{3L_r\alpha}{2} + \frac{iL_r\beta}{2} + i\beta L_s} \epsilon_1^2 \sqrt{1 - \epsilon_2^2}(e^{L_r\alpha} + e^{iL_r\beta}(-1 + \epsilon_1^2)(-1 + \epsilon_2^2)) \tag{A19}$$

$$K_{ya} = -\frac{1}{2} e^{-L_r\alpha + \frac{iL_r\beta}{2} + 2i\beta L_c + i\beta L_c} \sqrt{1 - \epsilon_1^2} \epsilon_2^2 (ie^{\frac{L_r\alpha}{2}} + e^{\frac{1}{2}i(L_r\beta + 2\beta L_s)}\epsilon_1^2 \sqrt{1 - \epsilon_2^2}) \tag{A20}$$

$$K_{ab} = -\frac{1}{2} e^{-\frac{1}{2}L_r(\alpha - i\beta) + i(2\beta L_c + \beta L_m)} \sqrt{1 - \epsilon_1^2} \epsilon_2^2 \tag{A21}$$

$$K_{bb} = e^{-L_r\alpha + iL_r\beta}(-1 + \epsilon_1^2)(-1 + \epsilon_2^2) \tag{A22}$$

$$K_{xb} = \frac{1}{2} e^{-L_r\alpha + \frac{iL_r\beta}{2} + 2i\beta L_c + i\beta L_m} \sqrt{1 - \epsilon_1^2} \epsilon_2^2 (-ie^{\frac{L_r\alpha}{2}} + e^{\frac{1}{2}i(L_r\beta + 2\beta L_s)}\epsilon_1^2 \sqrt{1 - \epsilon_2^2}) \tag{A23}$$

$$K_{yb} = -e^{-\frac{3L_r\alpha}{2} + \frac{iL_r\beta}{2} + i\beta L_s} \epsilon_1^2 \sqrt{1 - \epsilon_2^2}(e^{L_r\alpha} + e^{iL_r\beta}(-1 + \epsilon_1^2)(-1 + \epsilon_2^2)) \tag{A24}$$

$$K_{ax} = 0 \tag{A25}$$

$$K_{bx} = -\frac{1}{2} i e^{-\frac{1}{2}L_r(\alpha - i\beta) + i(2\beta L_c + \beta L_m)} \sqrt{1 - \epsilon_1^2} \epsilon_2^2 \tag{A26}$$

$$K_{xx} = e^{-L_r\alpha + iL_r\beta}(-1 + \epsilon_1^2)(-1 + \epsilon_2^2) \tag{A27}$$

$$K_{yx} = \frac{1}{2} e^{-L_r\alpha + \frac{iL_r\beta}{2} + 2i\beta L_c + i\beta L_c} \sqrt{1 - \epsilon_1^2} \epsilon_2^2 (ie^{\frac{L_r\alpha}{2}} + e^{\frac{1}{2}i(L_r\beta + 2\beta L_s)}\epsilon_1^2 \sqrt{1 - \epsilon_2^2}) \tag{A28}$$

$$K_{ay} = -\frac{1}{2} e^{-\frac{1}{2}L_r(\alpha - i\beta) + i(2\beta L_c + \beta L_m)} \sqrt{1 - \epsilon_1^2} \epsilon_2^2 \tag{A29}$$

$$K_{by} = 0 \tag{A30}$$

$$K_{xy} = \frac{1}{2} e^{-L_r\alpha + \frac{iL_r\beta}{2} + 2i\beta L_c + i\beta L_m} \sqrt{1 - \epsilon_1^2} \epsilon_2^2 (-ie^{\frac{L_r\alpha}{2}} + e^{\frac{1}{2}i(L_r\beta + 2\beta L_s)}\epsilon_1^2 \sqrt{1 - \epsilon_2^2}) \tag{A31}$$

$$K_{yy} = e^{-L_r\alpha + iL_r\beta}(-1 + \epsilon_1^2)(-1 + \epsilon_2^2) \tag{A32}$$

To analyze the behavior of the aforementioned coupled microring laser system and to simplify the calculation, we consider a lossless, resonating, non-Hermitian system with two identical cavities and ignore the phase noise introduced by the S-bend. Substituting the following parameters to Eq. (A16): $\alpha = 0$, $\beta L_r = 2\pi$, $\beta L_c = \pi/2$, $\beta L_m = \pi$, $\beta L_s = 2\pi$, $\epsilon_1 = 0.4$, $\epsilon_2 = 0.2$, and one can obtain the eigenvalues:

$$\widehat{\Lambda} = \begin{bmatrix} 0.885 \\ 0.805 - 0.068i \\ 0.805 + 0.068i \\ 0.731 \end{bmatrix} = \begin{bmatrix} 0.885 \\ 0.808 e^{-i0.084} \\ 0.808 e^{i0.084} \\ 0.731 \end{bmatrix} \tag{A33}$$

Here, the magnitudes of the elements represent the roundtrip amplification/damping of the fields and element arguments represent the roundtrip phase change of the fields. In a laser system, eigenmode with the highest amplification will reach lasing threshold first under pumping. The eigenvector that is related to eigenvalue $\Lambda_1 = 0.885$ represent the lasing mode in this condition. Its corresponding eigenvector is:

$$\begin{bmatrix} a \\ b \\ x \\ y \end{bmatrix} = \begin{bmatrix} 0.683 \\ 0.683 + 0.037i \\ -0.175 - 0.048i \\ -0.178 + 0.039i \end{bmatrix} \tag{A34}$$

In this case, the amplitudes of fields $a$ and $b$ are almost similar and significantly larger than those of $x$ and $y$ because of the S-bend. It is therefore, conceivable that the system indeed support a mode that is to a good approximation (especially after the gain nonlinearity is factored in) that is predicted from temporal coupled mode with a priori assumption that the two rings are unidirectional.

## B. Temporal coupled mode analysis

In this section, we show how a Hermitian antisymmetric imaginary coupling can be achieved by introducing a link with two directional couplers between two unidirectional ring resonators (Fig. 6).

When applying the antisymmetric imaginary coupling between two identical resonators (coupled to bus lines), in the tight-binding picture, one obtains the following time evolution equations[6]:

$$\begin{cases} i\frac{d}{dt}a = (-i\gamma_{in} - i\gamma_{ex})a + it_2 b - i\sqrt{2\gamma_{ex}}\varepsilon_{in} \\ i\frac{d}{dt}b = (-i\gamma_{in} - i\gamma_{ex})b - it_2 a \end{cases} \tag{B1}$$

where $a$ and $b$ are the complex field modal amplitudes in the two identical unidirectional ring resonators. $\gamma_{ex}$ is the coupling coefficient between the resonators and the input/output bus-waveguides. $\gamma_{in}$ represents the gain/loss in the two resonators and $t_2$ is associated with the tunneling rate. Let us assume an input of amplitude $\varepsilon_{in}$ at an angular frequency $\omega_0 + \delta\omega$ injected from the bus-waveguide to the resonator ① on the right, where $\omega_0$ is the resonance frequency of the resonators and $\delta\omega$ represents a shift form this value. The output fields propagating in the two bus-waveguides are given by $b_{out} = \sqrt{2\gamma_{ex}}b$ and $a_{out} = \varepsilon_{in} + \sqrt{2\gamma_{ex}}a$. As a result, the reflection and transmission coefficients of this system as obtained from the temporal coupled mode theory are:

$$R_{CM} = \frac{\sqrt{2\gamma_{ex}}b}{\varepsilon_{in}} = \frac{2t_2\gamma_{ex}}{t_2^2 + (i\delta\omega - \gamma_{ex} - \gamma_{in})^2} \tag{B2}$$

$$T_{CM} = \frac{\sqrt{2\gamma_{ex}}a + \varepsilon_{in}}{\varepsilon_{in}} = 1 + \frac{2\gamma_{ex}(i\delta\omega - \gamma_{in} - \gamma_{ex})}{t_2^2 + (i\delta\omega - \gamma_{ex} - \gamma_{in})^2} \tag{B3}$$

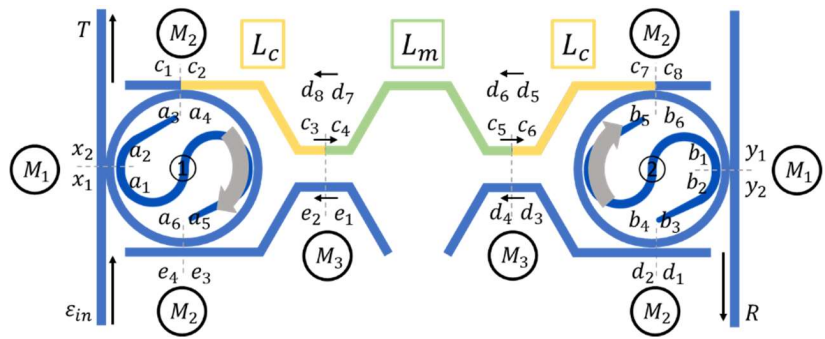

**FIG. 6.** Two microring resonators coupled through a link involving two directional couplers.

To self-consistently relate the above results (obtained from temporal coupled mode theory) to those expected from a continuous wave analysis, we introduce the transfer matrices[20,21] associated with the various elements involved in the links between the two resonators as shown in Fig. 6. In particular,

$$M_1 = \begin{pmatrix} \sqrt{1-\epsilon_1^2} & i\epsilon_1 \\ i\epsilon_1 & \sqrt{1-\epsilon_1^2} \end{pmatrix}, \quad M_2 = \begin{pmatrix} \sqrt{1-\epsilon_2^2} & i\epsilon_2 \\ i\epsilon_2 & \sqrt{1-\epsilon_2^2} \end{pmatrix} \tag{B4}$$

$$M_3 = \frac{1}{\sqrt{2}} \begin{pmatrix} 1 & i \\ i & 1 \end{pmatrix} \tag{B5}$$

where $\sqrt{1-\epsilon_j^2}$ and $i\epsilon_j$ represent the portion of the field that propagates through and cross the coupling regions, respectively. The $M_3$ matrices correspond to the two 3dB couplers, while the $M_1$ and $M_2$ matrices account for weak coupling to the waveguide buses, i.e. $\epsilon_j \ll 1$. The coupling between the bus-waveguides and the resonators are now written as:

$$\begin{pmatrix} x_2 \\ a_2 \end{pmatrix} = M_1 \begin{pmatrix} x_1 \\ a_1 \end{pmatrix}, \quad \begin{pmatrix} y_2 \\ b_2 \end{pmatrix} = M_1 \begin{pmatrix} y_1 \\ b_1 \end{pmatrix} \tag{B6}$$

and the coupling between the resonator and the links can be expressed via:

$$\begin{pmatrix} a_4 \\ c_2 \end{pmatrix} = M_2 \begin{pmatrix} a_3 \\ c_1 \end{pmatrix}, \quad \begin{pmatrix} b_4 \\ d_2 \end{pmatrix} = M_2 \begin{pmatrix} b_3 \\ d_1 \end{pmatrix} \tag{B7}$$

$$\begin{pmatrix} b_6 \\ c_8 \end{pmatrix} = M_2 \begin{pmatrix} b_5 \\ c_7 \end{pmatrix}, \quad \begin{pmatrix} a_6 \\ e_4 \end{pmatrix} = M_2 \begin{pmatrix} a_5 \\ e_3 \end{pmatrix} \tag{B8}$$

In addition, the field amplitudes at the 3dB directional couplers are

$$\begin{pmatrix} d_4 \\ d_6 \end{pmatrix} = M_3 \begin{pmatrix} d_3 \\ d_5 \end{pmatrix}, \quad \begin{pmatrix} d_8 \\ e_2 \end{pmatrix} = M_3 \begin{pmatrix} d_7 \\ e_1 \end{pmatrix} \tag{B9}$$

$$c_4 = \frac{1}{\sqrt{2}} c_3, \quad c_6 = \frac{1}{\sqrt{2}} c_5 \tag{B10}$$

We now assume that the effective propagation constant within the resonators is $k = 2\pi n_{\text{eff}}/\lambda_0$, where $n_{\text{eff}}$ is the effective refractive index of the lasing mode, $\lambda_0$ is the corresponding vacuum wavelength and $\alpha$ is the amplifying/damping rate per unit length. Given that $L_r$ represents the perimeter of each resonator, the field in resonator ① propagates according to

$$a_1 = e^{\frac{ikL_r}{4} - \frac{\alpha L_r}{4}} a_6, \quad a_3 = e^{\frac{ikL_r}{4} - \frac{\alpha L_r}{4}} a_2, \quad a_5 = e^{\frac{ikL_r}{2} - \frac{\alpha L_r}{2}} a_4. \tag{B11}$$

Similarly, the field components inside resonator ② obey

$$b_1 = e^{\frac{ikL_r}{4} - \frac{\alpha L_r}{4}} b_6, \quad b_3 = e^{\frac{ikL_r}{4} - \frac{\alpha L_r}{4}} b_2, \quad b_5 = e^{\frac{ikL_r}{2} - \frac{\alpha L_r}{2}} b_4. \tag{B12}$$

In addition, the length from the resonator to the directional coupler is $L_c$ and the length between two directional couplers is $L_m$, as shown in Fig. 6. By neglecting any amplification/loss in the link, the fields propagate according to:

$$c_3 = e^{ikL_c} \cdot c_2, \quad c_5 = e^{ikL_m} \cdot c_4, \quad c_7 = e^{ikL_c} \cdot c_6 \tag{B13}$$

$$d_3 = e^{ikL_c} \cdot d_2, \quad e_3 = e^{ikL_c} \cdot e_2 \tag{B14}$$

For lasing modes that operate near the resonant frequency in the unidirectional resonators, we assume that $e^{ikL_r} \to 1$ since $kL_r - 2m\pi = \delta k L_r \ll 1$, where $\delta k = (\delta\omega/c)n_{eff}$. Moreover, the loss in the resonators is negligible, i.e., $\alpha L_r \ll 1$, which results in $e^{-\alpha L_r} \to 1$). By expanding the exponential functions and square roots in the above equations, and keeping in mind that $\epsilon_j, \delta k L_r, \alpha L_r \ll 1$, we obtain the reflection coefficient of the system in the continuous wave picture:

$$R_{TM} = -\frac{2\epsilon_1^2 \epsilon_2^2}{P} \tag{B15}$$

where $P = i\sin(2kL_c + kL_m)\{[(2\alpha L_r - 2i\delta kL_r + \epsilon_1^2) + 2\epsilon_2^2]^2 - \epsilon_2^4\} - \cos(2kL_c + kL_m)\{[(2\alpha L_r - 2i\delta kL_r + \epsilon_1^2) + 2\epsilon_2^2]^2 + \epsilon_2^4\}$. To compare this reflection coefficient to that obtained from temporal coupled mode theory in Eq. (B2), we introduce the substitutions:

$$\epsilon_1^2 \to \frac{2\gamma_{ex}}{\text{FSR}}, \quad \epsilon_2^2 \to \frac{2t_2}{\text{FSR}}, \quad \alpha L_r = \frac{\gamma_{in}}{\text{FSR}}, \quad \delta kL_r = \frac{\delta\omega}{\text{FSR}}, \quad \text{FSR} = \frac{c}{n_g L_r} \tag{B16}$$

where $n_g$ is the group index. In this case, the reflection coefficient can be written as:

$$R_{TM} = -\frac{2t_2 \gamma_{ex}}{Q} \tag{B17}$$

where $Q = i\sin(2kL_c + kL_m)[(i\delta\omega - \gamma_{ex} - \gamma_{in} - 2t_2)^2 - t_2^2] - \cos(2kL_c + kL_m)[(i\delta\omega - \gamma_{ex} - \gamma_{in} - 2t_2)^2 + t_2^2]$. Notice that when $\phi = 2kL_c + kL_m = 2N\pi$ ($N =$ integer), the above reflection coefficient reduces to:

$$R_{TM} = \frac{2t_2 \gamma_{ex}}{(i\delta\omega - \gamma_{ex} - \gamma_{in} - 2t_2)^2 + t_2^2} \tag{B18}$$

which has exactly the same form as the reflection coefficient obtained from the temporal coupled mode theory (Eq. (B2)), once the coupling loss $-2t_2$ introduced by the link is also considered. In the case when $\phi = 2kL_c + kL_m = (2N-1)\pi$ ($N =$ integer) the directionality of the antisymmetric hopping is reversed, giving $t_2' = -t_2$.

The two eigenfrequencies of this system are given by:

$$\delta\omega_{1,2} = \pm t_2 - i\gamma_{ex} - i\gamma_{in} - 2it_2. \tag{B19}$$

Similarly, one can calculate the transmission coefficient of the system from the CW approach:

$$T_{TM} = 1 - \frac{2\epsilon_1^2[(2\alpha L_r - 2i\delta kL_r + \epsilon_1^2) + 2\epsilon_2^2]}{e^{2ik(2L_c + L_m)}\epsilon_2^4 + [(2\alpha L_r - 2i\delta kL_r + \epsilon_1^2) + 2\epsilon_2^2]^2}. \tag{B20}$$

The transmission coefficient can also be written as:

$$T_{TM} = 1 + \frac{2\gamma_{ex}[(i\delta\omega - \gamma_{ex} - \gamma_{in}) - 2t_2]}{e^{2i(2kL_c + kL_m)}t_2^2 + [(i\delta\omega - \gamma_{ex} - \gamma_{in}) - 2t_2]^2}. \tag{B21}$$

Again, when $\phi = 2kL_c + kL_m = N\pi$, where $N$ is an integer, the above transmission coefficient becomes:

$$T_{TM} = 1 + \frac{2\gamma_{ex}[(i\delta\omega - \gamma_{ex} - \gamma_{in}) - 2t_2]}{t_2^2 + [(i\delta\omega - \gamma_{ex} - \gamma_{in}) - 2t_2]^2} \tag{B22}$$

which has the same form as the transmission coefficient obtained from the temporal coupled mode theory (Eq. (B3)), apart from an additional term of $-2t_2$ introduced by the loss of the link.

As it is provided above, the reflection coefficient and transmission coefficient of the structure shown in Fig. 6 are given by Eqs. (B18) and (B22). This matches the forms of reflection and transmission coefficients of the system given by Eqs. (B2) and (B3). The structure in Fig. 6 hence behaves the same as such a system and Eq. (B16) is used to relate these parameters to the parameters used in the Hermitian Hamiltonian.